\documentclass[a4paper,11pt]{article}
\pdfoutput=1 

\usepackage{jheppub} 
\usepackage{xspace}
\usepackage[T1]{fontenc} 

\newcommand{\eemumu}{$e^+e^-\to\mu^+\mu^-$\xspace}
\newcommand{\LiteRed}{\texttt{LiteRed}\xspace}
\newcommand{\LiteRedII}{\texttt{LiteRed2}\xspace}
\newcommand{\Libra}{\texttt{Libra}\xspace}

\newcommand{\e}{{\ensuremath \epsilon}}

\newcommand{\jt}{\boldsymbol{J}}
\newcommand{\jc}{\boldsymbol{K}}

\title{\boldmath Two-loop master integrals for $e^{+}e^{-}\rightarrow\mu^+\mu^-$ process with account of electron mass.}
\author{Roman N. Lee}
\affiliation{Budker Institute of Nuclear Physics, 630090, Novosibirsk, Russia}
\emailAdd{r.n.lee@inp.nsk.su}

\abstract{We calculate a subset of two-loop master integrals relevant for the differential cross section of $e^+e^-\to \mu^+\mu^-$ process. We consider only those families for which the account of the electron mass $m$ is necessary. Our results have the form of the Frobenius series in $m$ with coefficients expressed via Goncharov's polylogarithms.}

\begin{document}
\maketitle
\flushbottom

\section{Introduction}
\label{sec:intro}

The process of muon pair production in electron-positron annihilation is probably the most fundamental QED process relevant for the electron-positron colliders. Consequently, it has a long history of investigation, starting from the calculation of the total Born cross section in Ref. \cite{BP56}. The next-to-leading order (NLO) corrections to the differential cross section were considered in Refs. \cite{Berends1983,Berends1973,Jadach1984}. Nowadays, the experimental precision has reached the point where the NNLO theoretical results for the differential cross section is needed. For this goal the calculation of the corresponding two-loop four-point master integrals is required. At present, the master integrals with zero electron mass have been already calculated in Ref.s  \cite{mastrolia2017master,di2018master,Lee:2019lno}. However, when inserted in the amplitude \cite{bonciani2022two}, they produce result which contains, in addition to the soft divergences, the collinear divergencies. While the former can be tamed by accounting for the soft photon contribution, the collinear divergences can not be treated in a similar way.\footnote{Note that there is an essential difference between the QCD processes and QED ones. While in the former there are always only colorless (zero color charge) in- and out-states, in the latter the experiments include charged particles. Therefore, in QCD, the collinear divergences disappear when the hard cross sections are integrated with parton distribution functions. But this is entirely due to the fact that hadrons are colorless.}
When the electron mass is taken into account, these collinear divergences turn into logarithms of the mass divided by some energy scale. Therefore, even though the electron mass is very small compared to any other scale in the whole kinematic region, one should take it into account when constructing physical observables.

\begin{figure}
	\includegraphics[width=\textwidth]{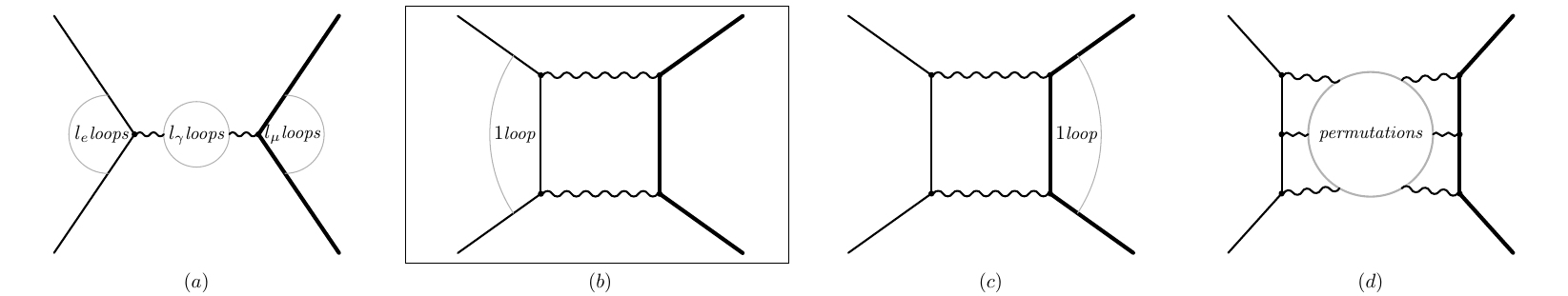}
	\caption{Gauge-invariant sets of diagrams that contribute to \eemumu at NNLO. First set contains one- and two-loop corrections to the electron and muon form factors and to the photon self-energy, $l_e+l_\mu+l_\gamma=2$. The diagrams considered in this paper correspond to the framed set $(b)$.}
	\label{fig:NNLOdiagrams}
\end{figure}

However, not all diagrams contribute to the collinear divergences. In Fig. \ref{fig:NNLOdiagrams} the sets of diagrams which contribute to the two-loop amplitude of \eemumu process are shown.

It can be shown that the collinear divergences appear only in the set $(a)$ with $l_e>0$ and in the set $(b)$, i.e., in the sets of diagrams where at least one photon line connects points on the electron line. Since the one- and two-loop corrections to the electron and muon form factors and to the photon self-energy, which contribute to set $(a)$, are already known, as well as the diagrams of the sets $(c)$ and $(d)$ at $m=0$, we are left with the problem of calculating diagrams $(b)$ at small but nonzero electron mass $m$. This is precisely the goal of the present paper.

Let us note that in this paper we do not consider the question of whether the master integrals for the diagrams of set $(b)$ can be evaluated exactly in $m$ in terms of polylogarithms or more complicated functions. Even if it were possible, the evaluation of a small-mass asymptotics in simpler form has its own value.

We use the approach based on the Frobenius expansion of the master integrals at small $m$ and the differential equations for the coefficients of this expansion. We use these differential equations to obtain the coefficients in terms of Goncharov's polylogarithms. Out approach is similar to that used in Ref. \cite{Lee:2024dbp} with two major differences. First, we manage to avoid using the complicated DRA approach by fixing boundary conditions using the asymptotics in several rather that in one kinematic limit. Second, in the present problem we initially have 4 scales $m,\, m_{\mu},\, s,\, t$, while in $e^+e^-\to 2\gamma$ process we have had only 3 scales $m,\, s,\, t$.

\section{Details of the calculation}

We consider the process
\begin{equation}
	e^-(p_1) + e^+(p_2) \longrightarrow \mu^-(q_1) + \mu^+(q_2)
\end{equation}
and introduce conventional invariants
\begin{equation}
	s=(p_1+p_2)^2=(q_1+q_2)^2,\quad t=(p_1-q_1)^2=(p_2-q_2)^2,\quad u=(p_1-q_2)^2=(p_2-q_1)^2\,.
\end{equation}
The momenta and invariants satisfy usual constraints
\begin{equation}
	p_1+p_2=q_1+q_2,\quad s+t+u =2m^2+2m_{\mu}^2,
\end{equation}
where $m^2=p_1^2=p_2^2$ and $m_\mu^2=q_1^2=q_2^2$ denote the squares of electron and muon masses, respectively. In what follows we put $m_\mu=1$ for convenience. We use dimensional regularization, $d=4-2\e$.

\begin{figure}
	\centering
	\includegraphics[width=0.7\textwidth]{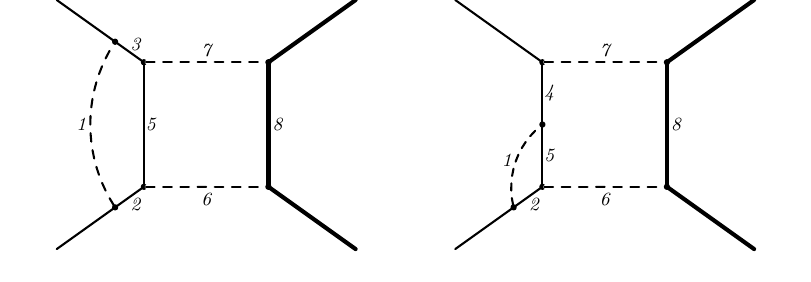}
	\label{fig:families}
	\caption{Two families belonging to the set $(b)$.}
\end{figure}

\begin{figure}[h]
	\includegraphics[width=\textwidth]{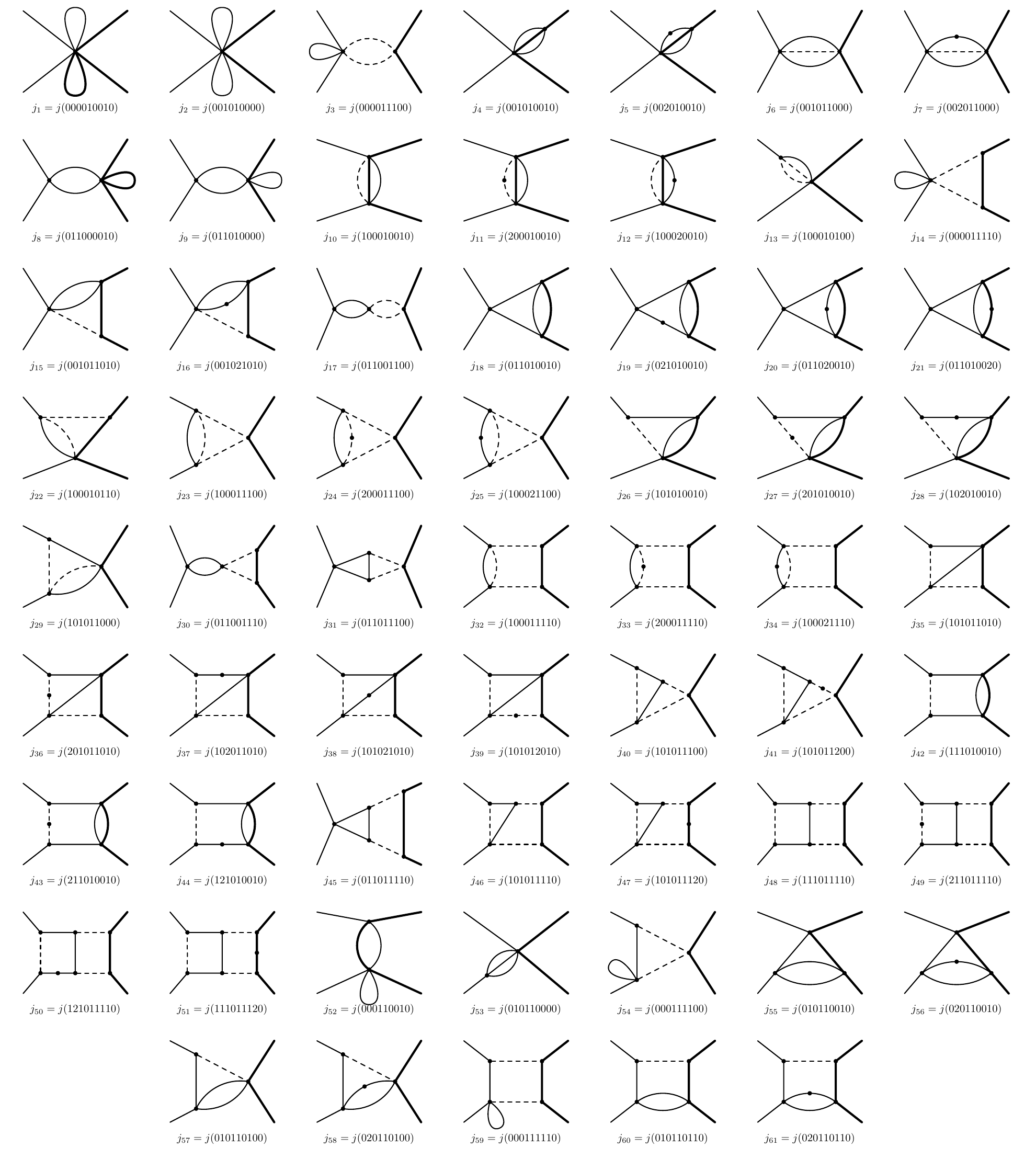}
	\label{fig:MIs}
	\caption{Master integrals and their graphs. Note that the numbering does not strictly follow the complexity of the integrals and the integrals $j_{52}-j_{61}$ are the additional master integrals which appear in the second family on Fig. \protect{\ref{fig:families}}.}
\end{figure}

The diagrams of the set $(b)$ on Fig. \ref{fig:NNLOdiagrams} are expressed in terms of the integrals of two big families depicted in Fig. \ref{fig:families}.
We define one \LiteRed basis, incorporating denominators of both diagrams:
\begin{equation}
	j(n_1,\ldots,n_9)=j(\boldsymbol{n})=\int\frac{d^dl_1d^dl_2}{\left(i\pi^{d/2}\right)^2}\prod_{k=1}^{9}D_k^{-n_k},
\end{equation}
where $n_9\leqslant0$, only one of $n_3$ and $n_4$ can be positive, and
\begin{gather}
	D_1=-l_1^2,\quad
	D_2=m^2-\left(p_1-l_1\right)^2,\quad
	D_3=m^2-\left(l_1+p_2\right)^2,\quad
	D_4=m^2-\left(l_2-p_1\right)^2,\nonumber\\
	D_5=m^2-\left(p_1-l_1-l_2\right)^2,\quad
	D_6=-l_2^2,\quad
	D_7=-\left(p_1+p_2-l_2\right)^2,\quad
	D_8=1-\left(l_2-q_1\right)^2,\nonumber\\
	D_9=l_1\cdot q_1\,.
\end{gather}
Using \LiteRedII \cite{Lee2013a,LeeLiteRed2}
we perform IBP reduction \emph{exactly} in $m$ (and other parameters) and reveal 61 master integrals, $\boldsymbol j= (j_1,\ldots,j_{61})^\intercal$ depicted in Fig. \ref{fig:MIs}. 

The differential equations for them have the form
\begin{gather}
	\partial_{m^2} \boldsymbol{j}=M_{m^2} \boldsymbol{j},\label{eq:dm}\\
	\partial_s \boldsymbol{j}=M_s \boldsymbol{j},\quad
	\partial_t \boldsymbol{j}=M_t \boldsymbol{j}\,,
\end{gather}
where $M_m,\ M_s,\ M_t$ are some rational matrices depending on $m,s,t$ and $\e$. For all transformations of the differential systems in this work we use \Libra \cite{Lee:2020zfb,LeeLibra} package.
Then we find the transformation $\boldsymbol{j}=T\jt$ which reduces the first system, Eq. \eqref{eq:dm}, to normalized fuchsian form at $m^2=0$. Thus we obtain the systems
\begin{gather}
	\partial_{m^2} \jt=\widetilde{M}_{m^2}\jt,\label{eq:dm1}\\
	\partial_s \jt=\widetilde{M}_s \jt,\quad
	\partial_t \jt=\widetilde{M}_t \jt\,,\label{eq:dst1}
\end{gather}
where $\widetilde{M}_{x}=T^{-1}({M}_{x}T-\partial_x T)$ and $\widetilde{M}_{m^2}=\frac{A_0}{m^2}+O(m^0)$. This allows us to apply the algorithm of Ref. \cite{Lee:2017qql} and to find the evolution operator (or fundamental matrix of solutions) $U=U(m^2,\underline{0})$ of Eq. \eqref{eq:dm1}, satisfying
\begin{equation}
	\partial_{m^2}U = \widetilde{M}_{m^2}U\,,\label{eq:dU}
\end{equation}
in the form
\begin{equation}
	U(m^2,\underline{0}) = \sum_{i=1}^{6} \left(m^2\right)^{\lambda_i} \sum_{n=0}^\infty\sum_{k=0}^{k_i} C(n+\lambda,k)(m^2)^n\frac{\ln^k \left(m^2\right)}{k!}.
	\label{eq:Useries}
\end{equation}
Here
\begin{align}
	\{\lambda_1\ldots \lambda_6\}&=\left\{0,\tfrac{1}{2}-2\e,-4 \e,-3 \e,-2 \e,-\e\right\},\nonumber\\
	\{k_1\ldots k_6\}&=\{0,0,0,0,1,1\},
\end{align}
and $C(n+\lambda,k)$ are some matrices with rational dependence on $s,t$ and $\e$.
In the present paper we find $C(n+\lambda,k)$ with $n\leqslant 6$, thus we find the expansion of $\jt$ up to $O(m^{13})$.

The inverse operator can be found in a similar form:
\begin{equation}
	U^{-1}(m^2,\underline{0}) = \sum_{i=1}^{6} \left(m^2\right)^{-\lambda_i} \sum_{n=0}^\infty\sum_{k=0}^{K_i} \widetilde{C}(n+\lambda,k)(m^2)^n\frac{\ln^k \left(m^2\right)}{k!}.\label{eq:Uiseries}
\end{equation}
Let us remark that the construction of the Frobenius expansions \eqref{eq:Useries} and \eqref{eq:Uiseries} was quite laborious and required extensive use of \texttt{Fermat} CAS \cite{Lewis}.
From the practical point of view it is important that, in order to construct \eqref{eq:Uiseries}, instead of inverting Eq. \eqref{eq:Useries}, we can use the fact that $\left(U^{-1}\right)^\intercal$ satisfies\footnote{This equation may be easily established by differentiating $1=U\,U^{-1}$.}
\begin{equation}
	\partial_{m^2}\left(U^{-1}\right)^\intercal = -\widetilde{M}_{m^2}^{\intercal}\left(U^{-1}\right)^\intercal
\end{equation}
and apply the same code that we used for the calculation of $U$ from Eq. \eqref{eq:dU}.

The specific solution has the form $\jt = U\jt_0$, where $\jt_0$ are the boundary constants depending on $s,t$, and $\e$. Using \Libra's procedure \texttt{GetLcs} \cite{Lee:2020zfb}, we relate  $\jt_0$ to specific $m\to 0$ asymptotic coefficients of the original integrals $\boldsymbol{j}$. Thus we obtain
\begin{equation}
	\jt = UL\boldsymbol{c},\label{eq:jviac}
\end{equation}
where $L$ is some matrix rational in $s,t,\e$ and $\boldsymbol{c}=\boldsymbol{c}(\e,s,t)$ is a column of asymptotic coefficients. To evaluate these coefficients, we construct differential equations for them with respect to $s$ and $t$. We use the same approach as in Ref. \cite{Lee:2024dbp}. Namely, we substitute \eqref{eq:jviac} into \eqref{eq:dst1} and obtain
\begin{equation}
	\partial_s \boldsymbol{c}=\overline{M}_s \boldsymbol{c},\quad
	\partial_t \boldsymbol{c}=\overline{M}_t \boldsymbol{c}\,,\label{eq:dst2}
\end{equation}
where
\begin{equation}
	\overline{M}_x=L^{-1}U^{-1}\left[\widetilde{M}_x UL -\partial_x(UL)\right],\qquad (x=s,t).
\end{equation}
Note that since $\boldsymbol{c}$ is independent of $m^2$, so are the matrices $\overline{M}_s$ and $\overline{M}_t$. Therefore, our chopped series results for $U$ and $U^{-1}$ were sufficient to find the \emph{exact form} of $\overline{M}_s$ and $\overline{M}_t$.

Then, using \Libra we find the transformation
\begin{equation}
	\boldsymbol{c}=\overline{T}\jc
\end{equation}
reducing the differential equations \eqref{eq:dst2} to $\e$-form. In order to do this, we were led to the necessity to pass from $s$ to $\beta$, the velocity of muons in c.m.s.. We also found it convenient to pass from $t$ to the scattering angle $\theta$. For the reference we present the corresponding relations:
\begin{equation}
	s=\frac{4}{1-\beta^2},\qquad t=-\frac{1+2 \beta  c+\beta ^2}{1-\beta ^2}\,,
\end{equation}
where $c=\cos\theta$.
The resulting differential system for the canonical basis $\jc$  can be represented in $d\log$-form:
\begin{equation}
	d\jc=\e dS \jc,
\end{equation}
where the matrix $dS$ has the form
\begin{equation}
	dS = \sum_{i=1}^{8} S_i d\ln p_i
\end{equation}
with the alphabet
\begin{gather}
	\{p_1,\ldots,p_8\}=\left\{1-c,1+c,1-\beta,\beta,1+\beta,1-\beta  c,1+\beta  c,1+2 \beta  c+\beta ^2\right\},
\end{gather}
and $S_i$ being the numerical matrices.

We fix the boundary conditions by evaluating the asymptotics of $\boldsymbol{c}$ in different limits. We find it possible to express required asymptotic coefficients in terms of hypergeometric ${}_{q+1}F_{q}$-functions which can be expanded in $\e$ via alternating multiple zeta values. More precisely, in the limit $\beta\to 0$ we find all but 4 required coefficients out of 61.
The missing information for the 3 out of these 4 constants was obtained by considering  the $\beta\to1$ asymptotics at $\theta=\pi/2$. 

However, one constant $\left[j_{41}\right]_{m^{-2-4\e}\beta^0}$, i.e., the coefficient in front of $m^{-2-4\e}\beta^0$ in the double asymptotics $m\to0,\beta\to0$ of $j_{41}=\!\!\!\!\!\raisebox{-3.5mm}{\includegraphics[width=15mm]{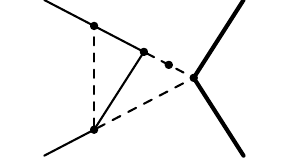}}$ required a special treatment. Namely, we had to consider a subset of 11 master integrals $\left\{j_2,j_3,j_6,j_7,j_{13},j_{23},j_{24},j_{25},j_{29},j_{40},j_{41}\right\}$ belonging to the hierarchy of integral $j_{41}$. Those are vertex-type integrals independent of muon mass and $t$. We have reduced the corresponding subsystem to $\e$-form treating the parameter $m$ \emph{exactly} and obtained boundary conditions from the asymptotics $s\to 4m^2$. This calculation allowed us to obtain the $\e$-expansion for the constant $\left[j_{41}\right]_{m^{-2-4\e}\beta^0}$ and also provided a number of nontrivial cross-checks for the integrals from the above list.

In order to obtain expressions for $\jc(\beta,\theta)$, we use the straight line path connecting the point $(0,\theta)$ and $(\beta,\theta)$ and represent the result in the form
\begin{align}
	\jc &= U_\beta \jc_0,\\
	U_\beta(\beta,\underline{0})&=\mathrm{Pexp}\Big[\e\intop_{\underline{0}}^\beta d S\Big]
	=\lim_{\beta_0\to0} \mathrm{Pexp}\Big[\e\intop_{\beta_0}^\beta d S\Big]\beta_0^{\e S_4}\,.
\end{align}
where $\jc_0 = \jc_0(\e)$ is a column of boundary constants. Using the constant transformations $T=T(\e)$ satisfying $[T,dS]=0$, we have secured the uniform transcendentality (UT) form of these constants. As a result, we obtain the UT $\e$-expansion of $\jc(\beta,\theta)$ in terms of the generalized polylogarithms $G(\boldsymbol{a}|\beta)$ with alphabet $\{0,\pm1,\pm(\cos\theta)^{-1},-e^{\pm i\theta}\}$.

The UT form of our results fo $\jc$ allowed us to examine possible linear relations between the elements of canonical basis. We searched for the constraints of the form
\begin{equation}
	\boldsymbol{C}\cdot \jc =0,\label{eq:constraints}
\end{equation}
where $\boldsymbol{C}$ is a vector of rational numbers. We have discovered $15$ constraints, which we have checked to be compatible with the differential equations and boundary constants. This has left us with $61-15=46$ independent entries of $\jc$.

\section{Results}

According to the considerations of the previous section, we present out results in the form
\begin{align}
	\boldsymbol{j}&=T\jt\label{eq:res1},\\
	\jt &= UL\overline{T}\jc  = \sum_{i=1}^{6} \sum_{n=0}^{o_{m}}\sum_{k=0}^{k_i} (m^2)^{\lambda_i+n}\frac{\ln^k \left(m^2\right)}{k!}\jt_{i,n,k} +O\left((m^2)^{o_{m}+1}\right),\label{eq:res2}\\
	\jc &= \sum_{n=0}^{o_{\e}}\e^{n}\sum_{\boldsymbol{a},d_{\boldsymbol{a}}\leqslant n} r_{n,\boldsymbol{a}}G(\boldsymbol{a}|\beta) +O\left(\e^{o_{\e}+1}\right)\label{eq:res3}
\end{align}
In the last relation $d_{\boldsymbol{a}}=d_{(a_1,\ldots,a_k)}=k$ is the number of entries in $\boldsymbol{a}$, and $r_{n,\boldsymbol{a}}$ are constant coefficients of transcendental weight $n-d_{\boldsymbol{a}}$ expressed via alternating Euler sums
$\zeta_{\boldsymbol{m}}$.
Consequently, we present our results in three files:
\begin{enumerate}
	\item \texttt{jtoJ.m} --- first relation \eqref{eq:res1} in the form of \textit{Mathematica} substitution rules.
	\item \texttt{JtoK$\langle o_m\rangle$.m}  --- second relation \eqref{eq:res2} in the form of  \textit{Mathematica} substitution rules.
	\item \texttt{KtoG$\langle o_\e\rangle$.m} --- third relation \eqref{eq:res3} in the form of substitution rules.
	\item \texttt{Numerics.nb} --- an example of using the obtained results for obtaining the numerical estimates of the integrals.
\end{enumerate}
Here $\langle o_m\rangle$ and $\langle o_\e\rangle$ in the file names denote the orders of expansions in $m^2$ and in $\e$, respectively. As the size of our complete results is rather large, we attach to the present paper  shallow expansions \texttt{JtoK2.m} and \texttt{KtoG4.m}, while the deeper expansions with $o_m=o_\e=6$ are available from the author by request. The attached results should be sufficient for our planned physical application. Note that in the files \texttt{JtoK$\langle o_m\rangle$.m} and \texttt{KtoG$\langle o_\e\rangle$.m} we present the results in terms of the reduced set $\{K_1,\ldots K_{46}\}$ which remains after the account of the constraints from Eq. \eqref{eq:constraints}.

\paragraph{Cross checks.}
As the calculation of the integral family considered in this paper was highly nontrivial, we have performed a thorough comparison of the presented results with the numerical results obtained using \texttt{Fiesta} \cite{Smirnov2022}. As our results concern the expansion of the master integrals near $m=0$, we have taken for comparison a small value of  $m$. However, since we have obtained rather deep expansion, up to $m^{13}$, it was expected that the comparison will show convincing agreement already for not so small values of $m$, e.g., for $m=1/2$, which we indeed observe. It worth noting that we were not able to obtain reliable \texttt{Fiesta}  result for the most complicated integrals $j_{48}-j_{51}$ at $d=4-2\e$. Instead, we used dimensional recurrence relation and performed numerical comparison for those integrals at $d=6-2\e$.

\section{Conclusion}

In the present paper we have considered the master integrals for the two-loop QED corrections to \eemumu process. We have concentrated on the two families which in the massless limit contribute to the collinear divergence of the process amplitude and thus require the account of the electron mass. We have calculated the master integrals of these two families in the form of the Frobenius expansion with respect to the electron mass with coefficients expressed via Goncharov's polylogarithms.

\acknowledgments I am grateful to V.S.Fadin for fruitful discussions.
This work has been supported by Russian Science Foundation (RSF) through grant No. 20-12-00205.

%
%

\bibliographystyle{JHEP}
\bibliography{asym}

\providecommand{\href}[2]{#2}\begingroup\raggedright\begin{thebibliography}{10}

\bibitem{BP56}
V.~B. Berestetskii and I.~I. Pomeranchuk, \emph{Formation of a $\mu$ - meson
  pair in positron annihilation}, {\emph{JETP} {\bfseries 2} (1956) 580}.

\bibitem{Berends1983}
F.~A. Berends, R.~Kleiss, S.~Jadach and Z.~Was, \emph{Qed radiative corrections
  to electron-positron annihilation into heavy fermions}, {\emph{Acta Phys.
  Pol., Series B;(Poland)} {\bfseries 14} (1983) }.

\bibitem{Berends1973}
F.~A. Berends, K.~Gaemers and R.~Gastmans, \emph{$\alpha$3-contribution to the
  angular asymmetry in e+ e-→ $\mu$+ $\mu$-}, {\emph{Nuclear Physics B}
  {\bfseries 63} (1973) 381}.

\bibitem{Jadach1984}
S.~Jadach and Z.~Was, \emph{Qed 0 ($\alpha$ 3) radiative corrections to the
  reaction e+ e-→ $\tau$+ $\tau$-including spin and mass effects},
  {\emph{Acta Physica Polonica. Series B} {\bfseries 15} (1984) 1151}.

\bibitem{mastrolia2017master}
P.~Mastrolia, M.~Passera, A.~Primo and U.~Schubert, \emph{Master integrals for
  the nnlo virtual corrections to $\mu$e scattering in qed: the planar graphs},
  {\emph{Journal of High Energy Physics} {\bfseries 2017} (2017) 1}.

\bibitem{di2018master}
S.~Di~Vita, S.~Laporta, P.~Mastrolia, A.~Primo and U.~Schubert, \emph{Master
  integrals for the nnlo virtual corrections to $\mu$e scattering in qed: the
  non-planar graphs}, {\emph{Journal of High Energy Physics} {\bfseries 2018}
  (2018) 1}.

\bibitem{Lee:2019lno}
R.~N. Lee and K.~T. Mingulov, \emph{{Master integrals for two-loop $C$-odd
  contribution to $e^+e^-\to \ell^+\ell^-$ process}},
  \href{https://arxiv.org/abs/1901.04441}{{\ttfamily 1901.04441}}.

\bibitem{bonciani2022two}
R.~Bonciani, A.~Broggio, S.~Di~Vita, A.~Ferroglia, M.~K. Mandal, P.~Mastrolia
  et~al., \emph{Two-loop four-fermion scattering amplitude in qed},
  {\emph{Physical Review Letters} {\bfseries 128} (2022) 022002}.

\bibitem{Lee:2024dbp}
R.~N. Lee and V.~A. Stotsky, \emph{{Master integrals for
  $e^{+}e^{-}\rightarrow2\gamma$ process at large energies and angles}},
  \href{https://arxiv.org/abs/2410.03336}{{\ttfamily 2410.03336}}.

\bibitem{Lee2013a}
R.~N. Lee, \emph{{LiteRed 1.4: a powerful tool for reduction of multiloop
  integrals}},  vol.~523, p.~012059, 2014,
  \href{https://arxiv.org/abs/1310.1145}{{\ttfamily 1310.1145}},
  \href{https://doi.org/10.1088/1742-6596/523/1/012059}{DOI}.

\bibitem{LeeLiteRed2}
R.~N. Lee, ``\texttt{LiteRed2}, essential update of \texttt{LiteRed} package.''
  \url{https://github.com/rnlg/LiteRed2}.

\bibitem{Lee:2020zfb}
R.~N. Lee, \emph{Libra: A package for transformation of differential systems
  for multiloop integrals},
  \href{https://doi.org/10.1016/j.cpc.2021.108058}{\emph{Comput. Phys. Commun.}
  {\bfseries 267} (2021) 108058}
  [\href{https://arxiv.org/abs/2012.00279}{{\ttfamily 2012.00279}}].

\bibitem{LeeLibra}
R.~N. Lee, ``\texttt{Libra}, package for transforming first-order linear
  differential systems.'' \url{https://github.com/rnlg/Libra}.

\bibitem{Lee:2017qql}
R.~N. Lee, A.~V. Smirnov and V.~A. Smirnov, \emph{{Solving differential
  equations for Feynman integrals by expansions near singular points}},
  \href{https://doi.org/10.1007/JHEP03(2018)008}{\emph{JHEP} {\bfseries 03}
  (2018) 008} [\href{https://arxiv.org/abs/1709.07525}{{\ttfamily
  1709.07525}}].

\bibitem{Lewis}
R.~Lewis, ``{Computer Algebra System Fermat}.''
  \url{http://home.bway.net/lewis/}.

\bibitem{Smirnov2022}
A.~V. Smirnov, N.~D. Shapurov and L.~I. Vysotsky, \emph{{FIESTA5: Numerical
  high-performance Feynman integral evaluation}},
  \href{https://doi.org/10.1016/j.cpc.2022.108386}{\emph{Comput. Phys. Commun.}
  {\bfseries 277} (2022) 108386}
  [\href{https://arxiv.org/abs/2110.11660}{{\ttfamily 2110.11660}}].

\end{thebibliography}\endgroup
\end{document}